\documentclass[twocolumn,prl,amssymb,amsmath,amsfonts]{revtex4}
\usepackage{graphicx}
\usepackage[usenames,dvipsnames]{xcolor}
\usepackage{mathrsfs}
\usepackage{bm}
\usepackage{wasysym}    
\usepackage{color}
\def\be{\begin{equation}}
\def\ee{\end{equation}}
\def\bc{\begin{center}}
\def\ec{\end{center}}
\newcommand{\blue}[1]{\textcolor{blue}{#1} }
\newcommand{\green}[1]{\textcolor{OliveGreen}{#1} }
\newcommand{\dagga}{{\phantom{\dagger}}}
\newcommand{\bea}{\begin{eqnarray}}
\newcommand{\eea}{\end{eqnarray}}
\usepackage[colorlinks,bookmarks=false,citecolor=blue,linkcolor=Green,urlcolor=NavyBlue]{hyperref}

\newcommand{\toadd}[1]{\textcolor{black}{#1}}

\begin{document}
\title{Bose glass transition and spin-wave localization for 2D bosons in a random potential}
\author{Juan Pablo \'Alvarez Z\'u\~niga}
\email{alvarez@irsamc.ups-tlse.fr}
\author{Nicolas Laflorencie}
\email{laflo@irsamc.ups-tlse.fr}
\affiliation{Laboratoire de Physique Th\'eorique, Universit\'e de Toulouse, UPS, (IRSAMC), F-31062 Toulouse, France}

\begin{abstract}
A spin-wave (SW) approach of the zero temperature superfluid --- insulator transition for two dimensional hard-core bosons in a random potential $\mu=\pm W$ is developed. While at the classical level there is no intervening phase between the Bose-condensed superfluid (SF) and the gapped disordered insulator, the introduction of quantum fluctuations leads to a much richer physics. Upon increasing the disorder strength $W$, the Bose-condensed fraction disappears first, {\it{before}} the SF. Then a gapless Bose-glass (BG) phase emerges over a finite region, until the insulator appears. Furthermore, in the strongly disordered SF regime, a mobility edge in the SW excitation spectrum is found at a finite frequency $\Omega_c$, decreasing with $W$, and presumably vanishing in the BG phase.
\end{abstract}
\maketitle
A correct understanding of the interplay between strong correlations and disorder is one of the most difficult questions in condensed matter physics~\cite{Evers08,Lagendijk09}.
While Anderson theory of localization~\cite{Anderson58} for single particle states is now a well-established paradigm to describe electronic transport in disordered environments, the equivalent bosonic problem of dirty superconductors or superfluids remains quite challenging~\cite{Fisher89, Weichman08}.
Despite numerous pionneer studies~\cite{Ma85-Ma86,Fisher89}, several questions remain open. For instance in 1D the universal character of the Luttinger exponent at the SF-BG transition has been recently questionned~\cite{Altman10,Zoran12,Vojta12}. For more realistic higher dimensional systems, relevant for disordered superconductors~\cite{Dubi07,Sacepe08-11}, quantum antiferromagnets~\cite{Hong10-Yu12}, or cold atoms~\cite{Sanchez10}, quantum Monte Carlo (QMC) approaches have considerably improved our understanding of the dirty boson problem all along the past twenty years~\cite{Krauth91,Runge92,Wallin94,Zhang95,Kisker97,Prokofiev04-Pollet09,Hitchcock06-Lin11,Baranger06}, but have also raised new issues regarding the universal value of some critical exponents~\cite{Hitchcock06-Lin11,Baranger06,Weichman07,Meier12}, and so far have only addressed ground-state properties. On the analytical side, important progresses have been made recently to go beyond mean-field (MF) theory~\cite{Ioffe10-Feigelman10,Benfatto12,Lemarie12,Monthus12a}. Although a naive MF is unable to find a localization transition, even at very strong disorder~\cite{Ma85-Ma86}, a quantum cavity approach on the Bethe~{\cite{Ioffe10-Feigelman10}} or the square lattice~{\cite{Monthus12a, Monthus12b, Lemarie12}} is able to capture such a transition. Nevertheless, several issues remain unsolved, in particular concerning finite frequency physics~{\cite{Ioffe10-Feigelman10, mueller2011magnetoresistance-gangopadhyay2012magnetoresistance, syzranov2012strong}}, and the outstanding question of many-body localization~{\cite{Basko06,Pal10,cuevas2012level}}.
\vskip .3cm

\begin{figure}[!t]
\centering
\includegraphics[width=\columnwidth,clip]{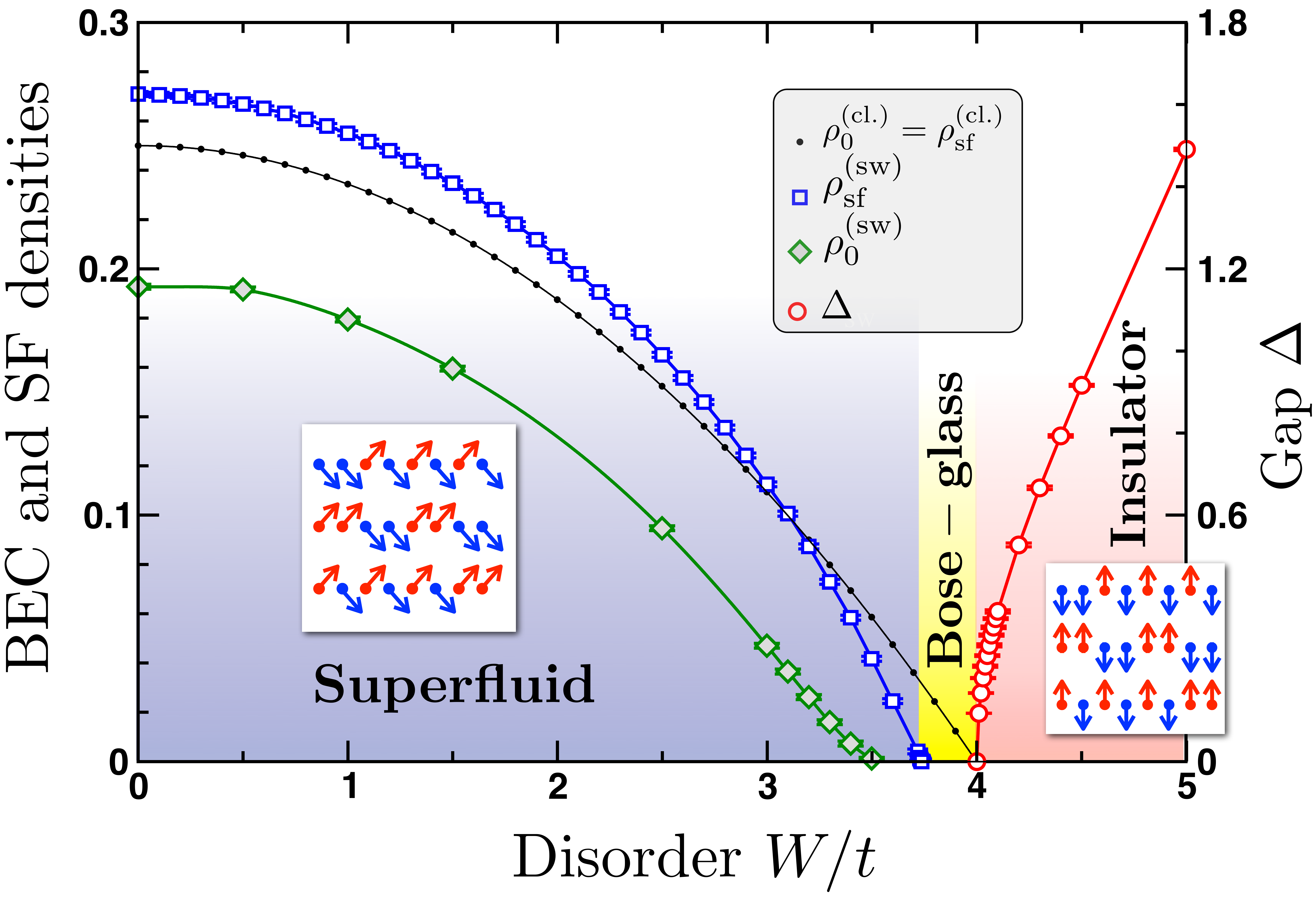}
\caption{(color online) Superfluid (SF) and Bose-Einstein condensate densities $\rho_{\rm sf}$ and $\rho_0$ plotted together with the gap $\Delta$, against the disorder strength $W/t$. The classical densities ($\bullet$) both vanish at the same point $W=4t$ whereas SW corrected quantities $\rho_{\rm sf}^{\rm(sw)}$ (${\blue{\square}}$) and $\rho_{0}^{\rm(sw)}$ (${\green{\diamond}}$) vanish at different points $W_{0}<W_{\rm sf}<4t$, leaving a finite window for an intervening gapless Bose glass before the gapped insulator. Insets depicts SF and insulating phases in the pseudo-spin representation. Disorder average was performed over several hundreds of disordered samples. The green line is a guide to the eyes.}
\label{fig:1}
\end{figure}

In this letter, we want to improve our understanding of the interplay between quantum fluctuations and disorder by addressing the spin-wave (SW) corrections for the Ma-Lee model in a disordered potential on the square lattice
\be
{\cal H}_{\rm b}=-t\sum_{\langle i j\rangle} \left(b_{i}^{\dagger}b_{j}^{\dagga} + b_{i}^{\dagga}b_{j}^{\dagger}\right) -W\sum_i  \epsilon_i n_i,
\label{eq:1}
\ee
which describes preformed Cooper pairs (hard-core bosons) hopping between nearest neighbor sites with a random chemical potential $W\epsilon_i$, where $\epsilon_i=\pm 1$ with probability $1/2$. 
In the disorder-free case (i.e. $\epsilon_i=1$ for instance), this well-known model~\cite{Bernardet02,Coletta12} displays two phases at $T=0$ : (i) a Bose-condensed superfluid regime for incommensurate filling $0<\langle n\rangle<1$ if $|W|<4t$, and (ii) a trivial insulator, filled (empty) with $\langle n\rangle =1$ ($\langle n\rangle =0$) for $W>4t$ ($W<-4t$). 
Using the Matsubara-Matsuda mapping~\cite{Matsubara56} of hard-core bosons onto pseudo-spin 1/2, Hamiltonian \eqref{eq:1} is exactly equivalent to a spin-$\frac{1}{2}$ XY model in a longitudinal field along the $z$ axis.
A mean-field description, where spin operators are treated as classical vectors with two angles $\theta_i$ and $\phi_i$, gives an energy ${\cal E}=-\frac{t}{2}\sum_{\langle i j\rangle} \sin\theta_i\sin\theta_j\cos(\phi_i-\phi_j) - \frac{W}{2} \sum_i \epsilon_i\cos\theta_i$, minimized by $\phi_i=$ constant and $\cos\theta_i=\epsilon_i W/(4t)$ if $W\le 4t$, meaning XY order for the spins (and superfluid Bose condensate for the bosons). If $W> 4t$ there is no XY order anymore: all spins point along the $z$ axis with $\cos\theta_i=\epsilon_i$ which, in the bosonic language, corresponds to a disordered insulator with local occupations $\langle n_i \rangle=(1+\epsilon_i)/2$ (= 0 or 1). In the XY regime, condensate and superfluid densities ($\rho_0$ and $\rho_{\rm sf}$) are both equal to $(\sin^2\theta_i)/4=\left[1-(W/4t)^2\right]/4$, vanishing at $W=4t$. Within such a classical description, a direct transition between SF and gapped phases is observed for $W>4t$, as visible in Fig.~\ref{fig:1}, with no intermediate localized regime, an artifact of MF theory.

However, when quantum fluctuations are introduced, the situation changes dramatically~\cite{Note1}. Before describing in more details our SW results, let us first briefly summarize our main conclusions. Here, we have studied square systems up to $64\times 64$ for several hundreds of disordered samples, which allowed us to get  infinite size extrapolations for various thermodynamic quantities such as the superfluid and the condensate densities $\rho_{\rm sf}$ and $\rho_0$. An intervening gapless Bose glass phase is unambiguously found between the superfluid and the gapped insulator. 
Properties of the SW excitation spectrum have also been studied, namely, the sound velocity and the inverse participation ratio (IPR) {of the SW excited states \cite{Ma93,Cea-tesi, alvarez-2013}}.
The localization of SW modes displays very interesting features vs frequency $\Omega$. We find a finite mobility edge $\Omega_c$, such that states with frequencies $\Omega<\Omega_c$ are extended and states at $\Omega>\Omega_c$ are localized. Upon increasing the disorder strength, $\Omega_c$ decreases and vanishes in the BG phase.\\
 
Let us now present in more details these results. SW corrections for hard-core bosons are treated in a straightforward way~\cite{Bernardet02,Coletta12}, first making a local rotation for the pseudo-spin operators, and then introducing Holstein-Primakoff bosons ($a,a^\dagger$). At the linear SW level, the hard-core bosons model \eqref{eq:1} reads ${\cal H}_{\rm b}={\cal E}+{\cal{H}}^{(2)}$, with 
\be
{\cal H}^{(2)}=-\frac{1}{2}\sum_{\langle i j \rangle}\Bigl[(t_{ij}a_i^{\dagga}a_j^{\dagger}
+\bar{t}_{ij}a_i^{\dagger}a_j^{\dagger})+{\rm{h.c.}} \Bigr] +\nu\sum_i n_i,
\label{eq:H2}
\ee
where $t_{ij}=t[1+\epsilon_i\epsilon_j({\bar{\nu}}/4t)^2]$, ${\overline{t}}_{ij}=t[\epsilon_i\epsilon_j ({\bar{\nu}}/4t)^2-1]$, with $\nu={\rm{max}}(W,4t)$ and ${\bar{\nu}}={\rm{min}}(W,4t)$. Because translational invariance is broken by the disorder, the quadratic bosonic Hamiltonian Eq.~\eqref{eq:H2} is diagonalized by a Bogoliubov transformation in real space which yields
\be
{\cal H}^{(2)}=\sum_{p=1}^{N}\Bigl[ \Omega_p(\alpha_p^\dagger\alpha_p^\dagga+\frac{1}{2})-\frac{\nu}{4}\Bigr].
\label{eq:H2diag}
\ee
$\Omega_p$ are the SW frequencies and ($\alpha^\dagga,\alpha^\dagger$) describe Bogoliubov quasi-particles. In the clean case $W=0$, the modes $p$ are labeled by the wave vectors ${\bf k}=(k_x,k_y)$ and the SW spectrum $\Omega_{\bf k}=2t\sqrt{4-2(\cos k_x+\cos k_y)}\approx 2t|{\bf k}|$ when $|{\bf k}|\to 0$, recovering the linear Bogoliubov spectrum with a "velocity of sound" $v_0=2t$. 

It is important to note that Bose-condensate and superfluid fractions are intrinsiqually different objects which are only equal in the simplest MF description; $^4$He being one of the best examples of a strongly correlated (non MF) bosonic system with $\rho_0/\rho_{\rm sf}\simeq 8\%$ at low temperature~\cite{Glyde00}. 
To go beyond MF, we want to compute the first SW corrections for the condensate and the superfluid response.
As discused in detail in Ref.~\onlinecite{Coletta12}, there are two ways for correctly computing $1/S$ corrections to a physical observable ${\cal O}$. One may evaluate its expectation value $\langle {\cal O}\rangle$
in the $1/S$-corrected ground-state, but this is not an easy task for our disordered problem. Perhaps more simply one can add a small symmetry-breaking term to the Hamiltonian of the form $\delta{\cal H}=-\Gamma {\cal O}$, compute the $1/S$-corrected energy and take the derivative with respect to the field in the limit $\Gamma\to 0$.  For instance, the condensate density 
$\rho_0=\frac{1}{N^2}\sum_{ij}\langle b^{\dagger}_{i}b^{\dagga}_{j}\rangle$, which 
is simply related in the pseudo-spin language to the transverse magnetization ($\rho_0=m_{xy}^2$ when $N\to \infty$) is obtained by adding a term $-\Gamma\sum_i S_i^x$ to the pseudo-spin  XY model. The SF density $\rho_{\rm sf}$ can be equally computed using the response of the system to twisted boundary conditions~\cite{Fisher73}, via the helicity modulus (or superfluid stiffness) $\Upsilon_{\rm sf}=\partial^2 E(\varphi)/\partial \varphi^2\bigr|_{\varphi=0}$, then simply related to the SF density by $\rho_{\rm sf}=\Upsilon_{\rm sf}/2t$. 

\begin{figure}[!b]
\centering
\includegraphics[width=\columnwidth,clip]{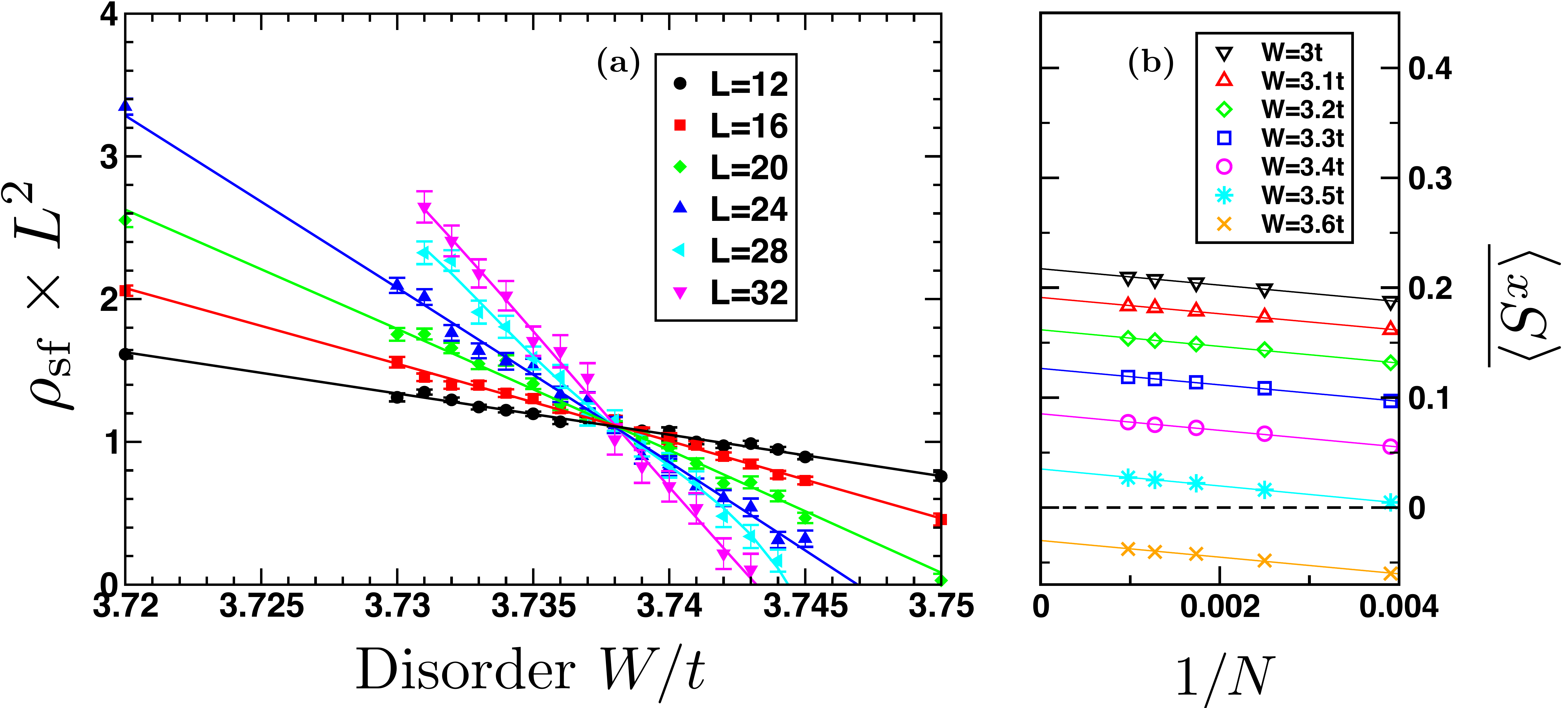}
\caption{(color online) (a) Crossing of the disorder average superfluid density $\rho_{\rm sf}\times L^{z}$ in the vicinity of the critical disorder where superfluidity disappears. Using $z=2$, a very convincing crossing is found for $W_{\rm sf}=3.738(1)$. (b) Transverse magnetization ${\overline{\langle S^x\rangle}}$, computed using a numerical derivative with respect to a small transverse field $\Gamma=t/100$ and averaged over several hundreds of samples, plotted against $1/N$ for various disorder strengths in the vicinity of $W_0=3.55(5)$.}
\label{fig:2}
\end{figure}

Numerical results for ${\overline{\langle S^x\rangle}}$ on lattices up to $32\times 32$ (averaged over several hundreds of disordered samples), are shown in panel (b) of Fig.~\ref{fig:2} versus $1/N$. There, we clearly see that when the disorder exceeds $W/t=3.5$, SW correction starts to become larger than the classical contribution, thus giving a negative magnetization which we interpret as a transition to a zero magnetization state. Finite size extrapolations to the thermodynamic limit [full lines in Fig.~\ref{fig:2}(b)] give the disorder average condensate density $\rho_0=({\overline{\langle S^x\rangle }})^2$~\cite{Note2}, plotted in Fig.~\ref{fig:1}. Such a behavior is not surprising as it is well-known that quantum fluctuations on top of the classical solution deplete the condensate 
mode. Here quantum fluctuations cooperates with disorder, leading to a monotonous destruction of Bose condensation, gradually increasing from $\sim 25\%$ depletion at $W=0$ up to $100\%$ at $W_0/t=3.55(5)$. 

More surprising is the behavior of the SF density $\rho_{\rm sf}$,
computed in the presence of a small twist angle $\varphi=10^{-2}$.
Infinite size extrapolations for $\rho_{\rm sf}$ are shown in Fig.~\ref{fig:1} (blue squares) where we see that contrary to the condensate, quantum fluctuations first enhance superfluidity for weak disorder, until $W/t=3$ where quantum and disorder effects start to cooperate and destroy the superfluid which finally disappears for a critical disorder $4>W_{\rm sf}/t=3.738(1) > W_0/t$. One can also test hyper-scaling at the 2D critical point where~\cite{Wallin94}
$\rho_{\rm sf}\sim L^{-z}$ is expected. As shown in Fig.~\ref{fig:2} (a) we check that the best crossing of $\rho_{\rm sf}\times L^z$ is obtained at $W_{\rm sf}/t=3.738(1)$ with a critical exponent $z=2.0(1)$, in a surprisingly good agreement with the expected $z=d$~\cite{Fisher89, Hitchcock06-Lin11}. 
{\toadd{A very careful QMC study is necessary~\cite{JPA13} in order to investigate whether such a scaling will survive to higher order corrections. Interestingly, condensate and superfluidity disappear for different values of the disorder, realizing a condensate-free superfluid~\cite{Laflorencie09-Laflo11-Kramer12}. While such a state of matter could in principle be stabilized in such a system, it is legitimate to wonder whether the window $W_{\rm sf}-W_0$ remains finite beyond linear SW corrections, a question perfectly suited to future QMC simulations~\cite{JPA13}. }}
In any 
case, we have demonstrated here that linear SW corrections can drive a bosonic state where both $\rho_0$ and $\rho_{\rm sf}$ are zero over a finite window $W\in\left[W_{\rm sf},4t\right]$, which is interpreted as a insulating Bose glass with a gapless excitation spectrum, as we discuss now.
\begin{figure}[!t]
\centering
\includegraphics[width=0.95\columnwidth,clip]{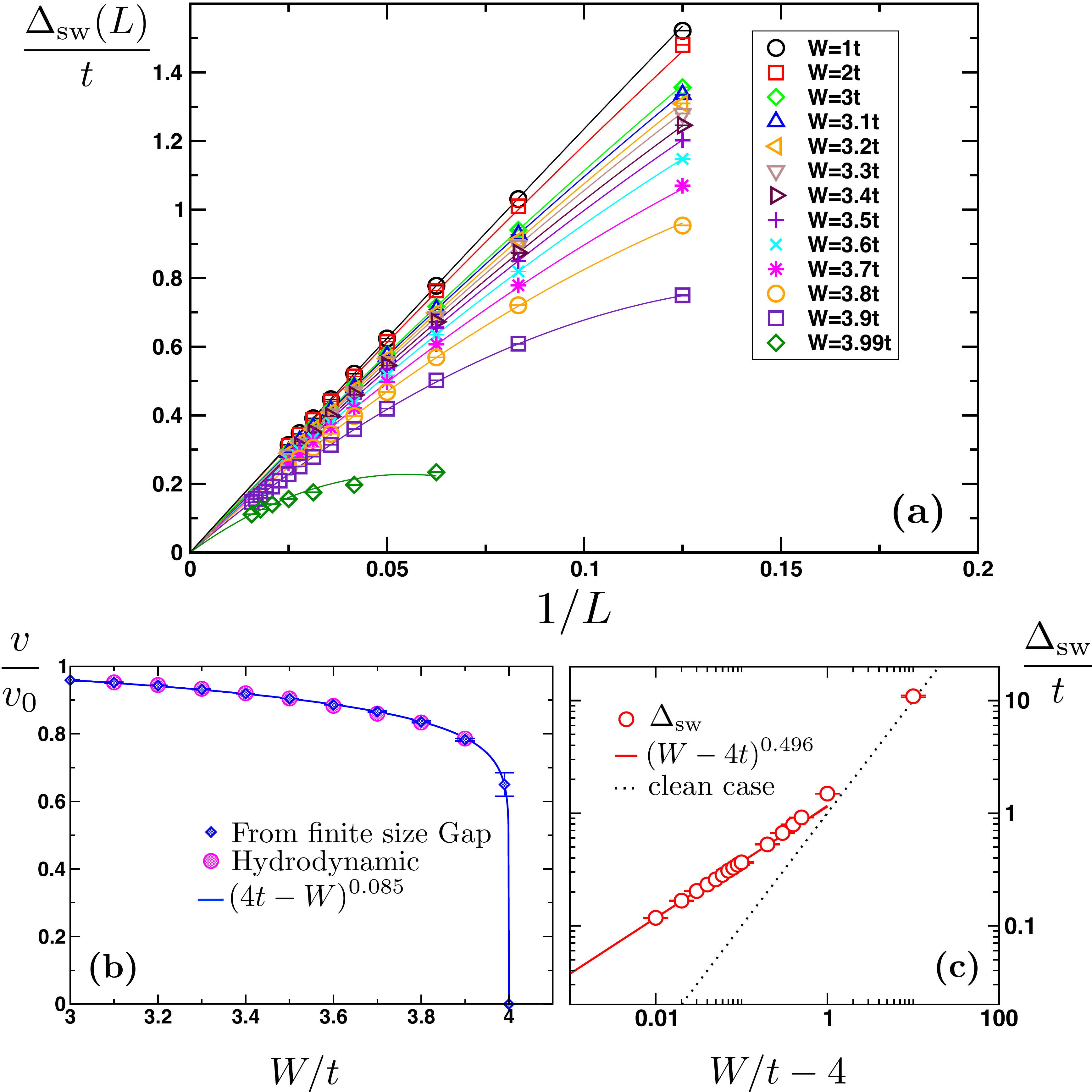}
\caption{(color online)  (a) Finite size SW gap plotted vs $1/L$ for various disorder strengths $W< 4t$ in the gapless regime. Full lines are quadratic fits of the form $\Delta(L)=2\pi v/L+b/L^2$, where $v$, the sound velocity, is displayed in panel (b) against $W/t$, together with the estimate from the classical hydrodynamic relation (see text). The full blue line is a power-law fit $\sim(4t-W)^{0.085}$. (c) Infinite size extrapolation of the SW gap (red circles) in the insulating regime $W>4t$.
The full red line is a power-law fit $\sim(W-4t)^{0.496}$, and the black dotted line is the clean case ($\mu=W>4t$) result: $\Delta=W-4t$.}
\label{fig:gaps}
\end{figure}

We first focus on the first excitation level above the Bogoliubov vaccum. We find the entire regime $0\le W/t\le 4$ to be gapless, with a zero mode $\Omega_0\simeq 0$, and a finite size gap to the first excitated state scaling in the limit $L\gg 1$, as $\Delta_{\rm sw}(L)\approx {2\pi v}/{L}$,
as visible in Fig.~\ref{fig:gaps} (a) for various values of the disorder $W$. The prefactor $v$ is identified with the velocity of sound (or SW velocity) and is shown in Fig.~\ref{fig:gaps} (b), rescaled by its zero-disorder value $v_0=2t$, versus $W/t$. In the same panel (b), the classical hydrodynamic relation for the velocity
$v=\sqrt{{\Upsilon_{\rm sf}}/{\kappa}}$,
is also plotted, $\Upsilon_{\rm sf}$ and $\kappa$ being the MF results for the helicity modulus and the compressibility. Both estimates for $v$ compare remarkably well.  Interestingly, the bottom of the SW spectrum is only weakly affected by the disorder and remains phonon-like (delocalized) over the entire gapless regime $W/t\in\left[0,4\right]$ with a finite velocity, almost disorder-independant, except very close to the insulating phase at $W/t=4$ where $v$ abruptely drops down~\cite{Note3}.
This finite velocity in the entire gapless regime is consistent with recent studies of Anderson localization of phonons in disordered solids~\cite{Monthus10,Amir12}.
Above $W=4t$ the zero mode disappears and a finite gap opens in the SW spectrum, as visible in Figs.~\ref{fig:1} and \ref{fig:gaps} (c). Interestingly, this gap does not scale linearly with $W-4t$ as in the clean case, but opens up more rapidly, presumably $\sim \sqrt{W-4t}$ and approaches the clean case only at large $W$.

\begin{figure*}[ht!]
\centering
\includegraphics[width=1.5\columnwidth,clip]{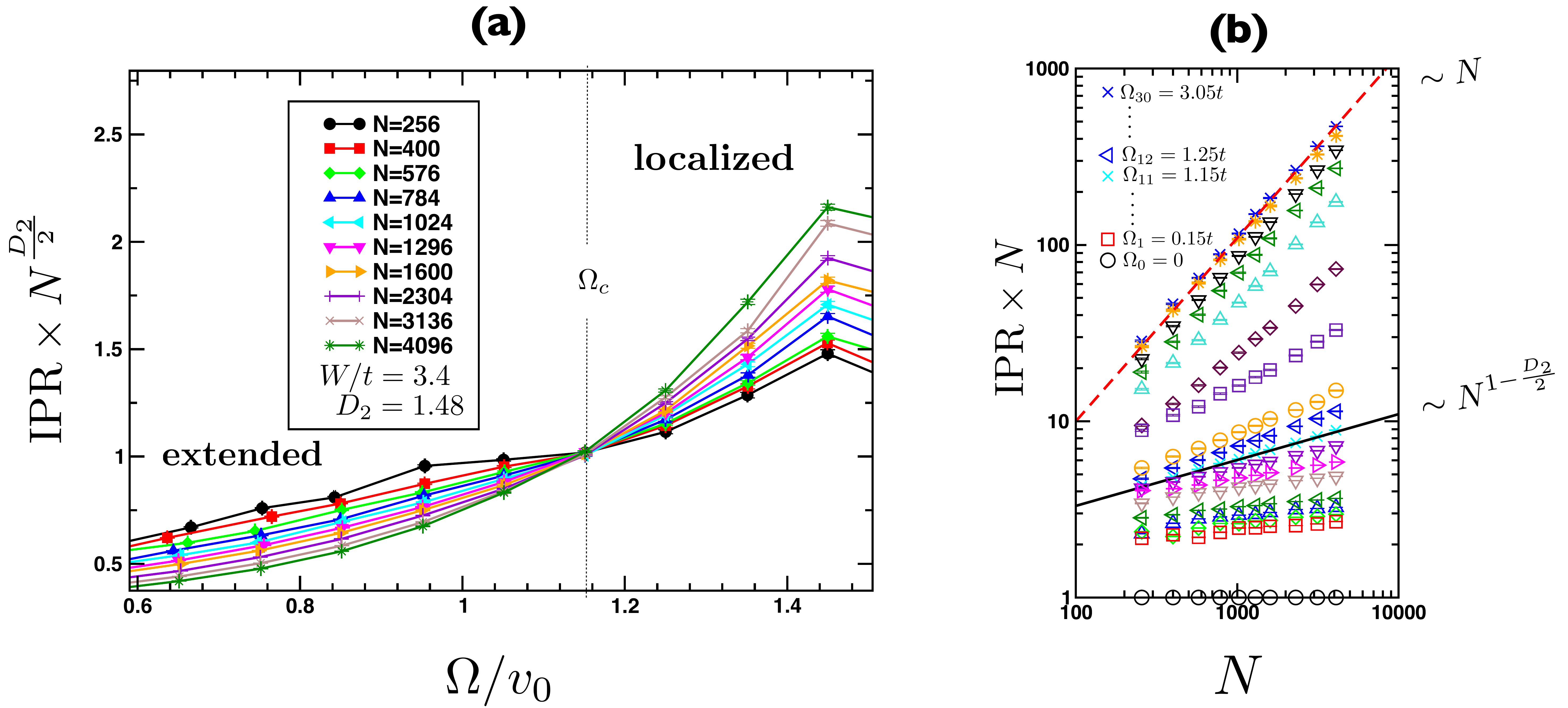}
\caption{(color online) {Inverse participation ratio IPR in the strongly disordered SF phase for $W=3.4t$; (a) Best crossing of IPR$\times N^{D_2/2}$ obtained with $N=256,\cdots, 4096$ at a mobility edge $\Omega_c/v_0\simeq 1.15$ with a fractal dimension $D_2=1.48$. (b) IPR$\times N$ plotted vs $N$ for different frequencies $\Omega_i=\Omega/v_0$ (different symbols). 
The dashed red line $\sim N$ shows the fully localized case when $N\gg\xi^2$ and the full black line is the critical scaling $\sim N^{1-D_2/2}$ at the mobility edge.}
}\label{fig:4}
\end{figure*}

{Following~\cite{Cea-tesi}, we have investigated the localization properties of the entire SW Bogoliubov excitation spectrum. Here we shall just mention the main results of this study which will be described in details in a longer article~\cite{alvarez-2013}. In Ref.~\onlinecite{Cea-tesi}, it has been observed that the localization properties of the SW excited states depend crucially on the frequency in a way similar to the Anderson localization of phonons~\cite{Monthus10}. Here, we have analyzed this effect by considering the inverse participation ratio} (IPR), defined for each (normalized) state $|p\rangle =\sum_i a_{i}^{p}|i\rangle$, where $i$ are lattice sites, by
${\rm{IPR}}_p={\sum_{i=1}^{N}|a_{i}^{p}|^{4}}$.
For delocalized modes IPR $\sim 1/N$ whereas localized states display a finite IPR $\sim 1/\xi^2$, where $\xi$ is the localization length. Since SW spectra are discrete for finite size systems, in particular at low energy, we define disorder average IPRs over finite slices of frequencies centered around $\Omega$:
\be
{\rm IPR}(\Omega)=\frac{\sum_p\Theta(\Omega_p,\Omega\pm\delta\Omega){\rm IPR}_p}{\sum_p \Theta(\Omega_p,\Omega\pm\delta\Omega)},
\ee
where $\Theta(\Omega_p,\Omega\pm\delta\Omega)=1$ if $\Omega-\delta\Omega\le\Omega_p\le\Omega+\delta\Omega$, and 0 otherwise, with $\delta\Omega/v_0=1/20$ in the following.{While for weak disorder $W/t<2$ all the excited states are found delocalized, similarly to the clean case where} the coefficients are simply the Fourier modes $a_i^p=\exp({\rm i}{\bf{k}}_p\cdot{\bf r}_i)/\sqrt{N}$, thus giving for all frequencies IPR$(\Omega)\times N ={\cal{O}}(1)${, the case of strongly disordered SF} appears much more interesting, as visible in Fig.~\ref{fig:4} which shows representative results for $W/t=3.4$. At low energy the modes are delocalized, but the situation changes dramatically above a certain threshold frequency $\Omega_c$ - the mobility edge - where ${\rm IPR}(\Omega)\times N$ starts to increase linearly with $N${, a characteristic signature of localization}. At the mobility edge{, as in the case of the Anderson transition \cite{Evers08, Monthus10},} the IPR is found to display an 
anomalous scaling ${\rm IPR}(\Omega_c)\propto N^{-D_2/2}$ with a fractal dimension $D_2\simeq 1.48<2$. 
This is well visible in the panel (a) of Fig.~\ref{fig:4} where the best crossing of IPR$(\Omega)\times N^{D_2/2}$ has been obtained for $D_2=1.48$. 
For other disorder strengths {\toadd{(as well as for other types of disorder~\cite{alvarez-2013})}}, the same fractal exponent has been found to obtain the best crossing curves, separating extended modes at $\Omega<\Omega_c$ from localized ones at $\Omega>\Omega_c$ (see ~\cite{SM}).
\begin{figure}[hb!]
\centering
\includegraphics[width=0.95\columnwidth,clip]{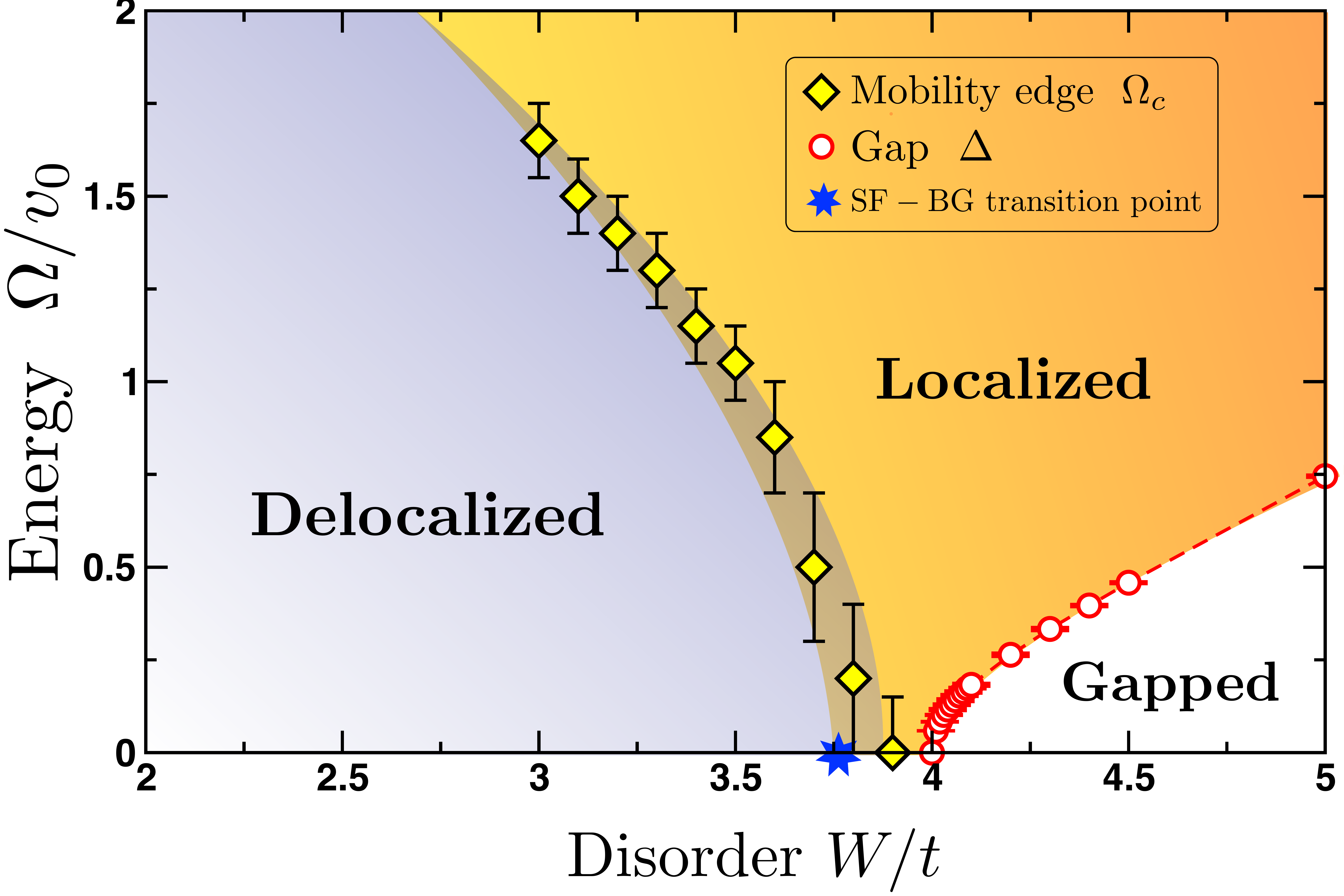}
\caption{(color online) Energy of the SW excitations $\Omega$ (in units of $v_0$) as a function of $W/t$.
All states are extended (delocalized) below the mobility edge $\Omega_c$ and localized above. The shaded area represents the localized-delocalized boundary with quite large error bars close to the SF-BG transition point where we expect $\Omega_c\to 0$. In the gapped insulating side, there is no state below the gap $\Delta$, and all excitations above are localized, and connected to other localized excitations.}
\label{fig:5}
\end{figure}

The evolution of the mobility edge $\Omega_c$ against increasing disorder is shown in Fig.~\ref{fig:5} where we see that $\Omega_c\to 0$ when the BG phase is approached. While the localization transition point is easily identified in Fig.~\ref{fig:4} for $W/t=3.4$, closer to the SF-BG boundary error bars for $\Omega_c$ get bigger. Indeed, it becomes more difficult to correctly estimate the localization transition on finite size systems for $W/t > 3.7$ where the crossing point displays a sizable drift towards smaller frequencies when $N$ increases. {\toadd{Nevertheless, our data are consistent with a zero frequency mobility edge in the BG state (see ~\cite{SM}), supporting the fact that the BG phase is localized for all $\Omega>0$.}} The phase diagram energy - disorder in Fig.~\ref{fig:5} displays 3 different regimes: (i) delocalized excitations in the SF regime below a finite mobility edge $\Omega_c$; (ii) absence of modes below a finite gap $\Delta$ for $W/t>4$; (iii) localized excitations above $\Omega_c$ or $\Delta$.
Finally one can mention that, contrary to Refs.~{\onlinecite{Mueller09,Ioffe10-Feigelman10}}, {inside the insulating phase, we do not find any mobility edge from localized excited states at small frequency to extended states at large frequencies}. Conversely, our results support the idea that superfluidity emerges out of the localized BG phase by a delocalization at $\Omega>0$, in agreement with Refs.~{\onlinecite{mueller2011magnetoresistance-gangopadhyay2012magnetoresistance,Cea-tesi}}

{\toadd{To conclude, we have shown that linear spin-wave corrections are able to capture the localization of 2D hard-core bosons in a random potential.
At $1/S$ order, an interesting condensate-free superfluid state is found, before entering in the disordered gapless Bose glass. The spin-wave excitation spectrum displays very interesting features, with a mobility edge at finite frequency above the superfluid phase, vanishing in the Bose glass.}}

\begin{acknowledgements}
{We thank T. Cea and C. Castellani for suggesting the study of localization of the SW excited states and for communicating their results prior to publication. We also thank G. Lemari\'e for his help in the analysis of the IPR scaling.} Part of this work has been supported by the French ANR project Quapris, and by the Labex NEXT.
\end{acknowledgements}

\end{document}